\documentclass[iop,twocolumn]{emulateapj}

\usepackage{soul}
\usepackage{placeins}
\usepackage{graphics}
\usepackage{color} 
\usepackage{verbatim} 
\usepackage{booktabs}
\usepackage{amsmath}

\def\2pr{^{\prime \prime}} 
 
\def\greatsim{\mathrel{\raise.3ex\hbox{$>$\kern-.75em\lower1ex\hbox{$\sim$}}}} 
\def\lesssim{\mathrel{\raise.3ex\hbox{$<$\kern-.75em\lower1ex\hbox{$\sim$}}}} 
\def\gs{\mathrel{\raise0.27ex\hbox{$>$}\kern-0.70em % Greater/squiggles 
\lower0.71ex\hbox{{$\scriptstyle \sim$}}}} 
\def\ls{\mathrel{\raise0.27ex\hbox{$<$}\kern-0.70em % Less than/squiggles 
\lower0.71ex\hbox{{$\scriptstyle \sim$}}}}

\slugcomment{Submitted to ApJ} 
 
\shorttitle{Spectral Variability of High-Redshift Quasars} 
\shortauthors{Dyer et al.}

\begin{document} 
 
\title{The Spectroscopic Signature of Variability in High-Redshift Quasars} 

\author{
Jamie~C.~Dyer\altaffilmark{1},
Kyle~S.~Dawson\altaffilmark{1},
H{\'e}lion~du~Mas~des~Bourboux\altaffilmark{1},
M.~Vivek\altaffilmark{1,2},
Dmitry~Bizyaev\altaffilmark{3,4},
Audrey~Oravetz\altaffilmark{3},
Kaike~Pan\altaffilmark{3}, and
Donald~P.~Schneider\altaffilmark{2,5}
}

\altaffiltext{1}{
Department of Physics and Astronomy,
University of Utah, Salt Lake City, UT 84112, USA.
}
\altaffiltext{2}{
Department of Astronomy and Astrophysics, 525 Davey Laboratory,
The Pennsylvania State University, University Park, PA 16802, USA.
}
\altaffiltext{3}{Apache Point Observatory and New Mexico State
University, P.O. Box 59, Sunspot, NM, 88349-0059, USA}
\altaffiltext{4}{Sternberg Astronomical Institute, Moscow State
University, Moscow, Russia}
\altaffiltext{5}{
Institute for Gravitation and the Cosmos,
The Pennsylvania State University, University Park, PA 16802, USA.
}

% \email{jamjam.dyer@gmail.com} 
 
\begin{abstract}

Using 16,421 spectra from a sample of 340 quasars ($1.62<z<3.30$) from the SDSS Reverberation Mapping Project, we present an analysis of quasar spectral variability.  We confirm the intrinsic Baldwin Effect and brighter-means-bluer trends in which emission line strength and color are associated with changes in luminosity. We then create a composite differential spectrum that captures the wavelength dependence of quasar variability as a function of change in luminosity. When using a bandpass around 1740 \AA\ to describe the luminosity, the differential spectrum follows a power law at longer wavelengths that breaks blueward of 1700 \AA.  However, the shape of the differential spectrum, the location of the power law break, and the slope of the intrinsic Baldwin Effect all vary with the choice of bandpass used to define the change in quasar luminosity. We propose that the observed behavior can be explained by inhomogeneous accretion or slim accretion disk models where delays in the reprocessing of incident light on the accretion disk cause local deviations in temperature from the thin disk model.  Finally, we quantify the effects on cosmology studies due to the variations in the quasar spectrum in the Lyman-$\alpha$ forest wavelength range. Using the observed spectroscopic signatures to predict the quasar continuum over the interval $1040 < \lambda < 1200$ \AA, we find that the derived spectral templates can reduce the uncertainty of the Lyman-$\alpha$ forest continuum level in individual epochs from 17.2\% to 7.7\%, near the level where systematic errors in SDSS flux calibration are expected to dominate.

\end{abstract}
 

\section{\textbf{Introduction}}\label{sec:intro}

Quasar spectra are identified by the common signatures of broad and narrow emission lines and an approximate power law continuum. The thin disk model \citep{thindisk} explains these characteristics in terms of an optically thick, geometrically thin accretion disk modeled as homogeneous, concentric blackbody rings. Broadband X-ray, UV, and optical emission are radiated from the accretion disk as a result of high temperatures (explicitly for the thin disk model $T \propto r^{-3/4}$). The high frequency photons are absorbed in the Broad Line Region (BLR) and Narrow Line Region (NLR) farther from the central engine, thus triggering ionization, excitation, and re-emission.

The thin disk model provides a general picture of quasar structure, but it is unable to explain prevalent trends in quasar spectra. One such trend is the location of an apparent break in quasars' spectral index around 1000~\AA, indicative of a maximum accretion disk temperature around 50,000~K \citep[e.g.][]{Stevans_Shull}. Another such correlation is the Baldwin Effect \citep[BE;][]{BEnature}. The BE refers to an empirical relationship in which quasars' continuum or bolometric luminosity ($L_{\mathrm{bol}}$) is anti-correlated with \textsc{C~iv} equivalent width ($W_\lambda$). The BE has been observed across large samples of quasars for many different emission lines \citep[e.g.][]{BEMgII,BEUVFe,BECroom,BEDietrich,XuBianMBH}. 

The strictly empirical nature of the BE makes it difficult to infer whether luminosity is the sole, fundamental parameter driving variations in quasar flux. Other possible drivers could be the mass of the black hole ($M_\mathrm{BH}$), its accretion rate ($\dot{M}_\mathrm{BH}$), Eddington Ratio ($R_{\mathrm{Edd}}$), color profile or spectral index ($\alpha_\lambda$), or orientation. Physical and systematic correlations between many of these parameters complicate the mapping of empirical observations onto physical mechanisms. For example, luminosity of quasars is known to increase with $M_\mathrm{BH}$ and $\dot{M}_\mathrm{BH}$, and the evolution of quasar growth rate correlates both of these parameters with redshift. 

\citet{Treyper} recently used composite differential spectra to explore diversity with respect to $L_{\mathrm{bol}}$, $z$, and $\alpha_\lambda$ across the same large sample used for baryon acoustic oscillation (BAO) studies with the Lyman-$\alpha$ forest \citep[$2.1 \le z \le 3.5$;][]{Bautista, Helion}. Differential spectra approximate partial derivatives of flux across all wavelengths with respect to an observable parameter. After careful mitigation of selection bias, \citet{Treyper} show that spectral diversity with respect to redshift has nearly the same signature as spectral diversity with respect to luminosity. Contrary to the association of luminosity with diversity in the typical BE, the signature appears to capture the redshift evolution of the global physical properties that drive quasar diversity. 

Observations from time-varying measurements of single quasars further challenge the thin disk model. The most blatant evidence in contention with the thin disk model is a recent finding that extreme UV variability occurs too rapidly for the gradual evolution expected for homogeneous accretion disk structure \citep{Punsly}. A unique Baldwin Effect is also observed as a single object varies in luminosity, showing a steeper relationship than the quasar-to-quasar, or ``global,'' BE described above \citep[e.g.][]{Kinney,Wilhitecivline}. Multi-epoch investigation of quasars has additionally revealed that quasars become increasingly blue as they brighten \citep[e.g.][]{Wilhite_wavelength, Pereyra2006, photometric_analog}. \citet{Pereyra2006} compared a composite spectrum derived from the residuals between the brightest and faintest epochs from \citet{Wilhite_wavelength} to a synthetic residual spectrum that was simulated from the thin disk model. They reported that the observed bluer-means-brighter effect is reproducible under the thin disk model if each object had sufficient time to adjust to a new mean accretion rate between observations. 

\citet{Ruan2014} created a relative variability spectrum (qualitatively similar to a residual spectrum) from a larger sample of quasars than was investigated in \citet{Pereyra2006}. They found the observed variability follows a power law inconsistent with changes in the overall accretion rate as predicted by the thin disk model. \citet{Ruan2014} compared their relative variability spectrum to synthetic ones generated under a model where the accretion disk is comprised of many independently varying zones which undergo large, ephemeral temperature fluctuations \citep[the inhomogeneous disk model;][]{DexterAgol}. The simulated relative variability spectra are consistent with the observed data over the full wavelength range (1300~-- 6000~\AA) for a given number of zones in the disk and amplitude of temperature fluctuations. The simulations showed, however, that there would be a break in the relative variability power law if fewer zones undergoing smaller temperature fluctuations were present in the disk. This break could exist in the observed relative variability spectrum in \citet{Ruan2014}, but be concealed by the limited UV coverage of their data.

\citet{Ruan2014}'s constraint on inhomogeneities in the accretion disk is consistent with results from independent micro-lensing studies on the size of accretion disks (see \citet{DexterAgol} and references therein). However, other studies raise questions about the inhomogeneous disk model. Through multi-epoch imaging, \citet{Kokubo2015} found intrinsic scatter across different photometric band-passes to have tighter correlations than would be possible if zones in the disk truly fluctuated independently. The inhomogeneous disk model of \citet{DexterAgol} also does not address the break in quasars' spectral index observed around 1000 \AA, nor the timescales of variability observed in \citet{Punsly}. The 1000 \AA\ break in the spectral index arguably requires a model where the inner region of the accretion disk is truncated due to outflowing, line-driven winds that strengthen with increasing Eddington efficiency \citep{leighly05,Bonning2013,LaorAndDavis,LuoBrandt2015}. At near-Eddington efficiencies, such advection can lead to accretion disk structure that is ``puffed-up'' in the vertical direction at small radii \citep[the slim disk model;][]{SlimDisk_old}. More multi-epoch data are needed to isolate the number and type of parameters responsible for quasar variability and in turn ultimately determine the correct accretion disk model.  

A deeper understanding of quasar astrophysics will enhance cosmology studies with quasars’ Lyman-$\alpha$ forests. Existing analyses on the one-dimensional power spectrum \citep[e.g.,][]{1dforneutrinos} and three-dimensional correlation function \citep[e.g.][]{Bautista, Helion} address quasar diversity by assuming linear deviations from an average quasar spectrum.  The continuum model is fit directly to the data in each quasar's Lyman-$\alpha$ forest.  The model not only captures the linear deviation of the quasar continuum from the average, but also systematically biases the mean and first moment of the density field to zero for each line of sight. Although studies indicate no significant bias in the measurement of the BAO distance scale, the diversity of continuum spectra introduced through quasar variability must introduce higher order modes that increase the statistical uncertainty in the derived distance scale.

If one were able to incorporate a full model for quasar diversity without requiring constraints from the observed Lyman-$\alpha$ forest, one could make unbiased estimates of the unabsorbed continuum.  Such a significant step would reduce the uncertainty in measurements of the cosmic distance scale with existing data, provide a mechanism to perform absolute measurements of neutral hydrogen density fluctuations, and enable new cosmological analyses using the larger scale modes of the Lyman-$\alpha$ forest density field.

Growing interest in reverberation mapping \citep[RM, e.g.,][]{blandford82a,peterson93a}, a process which aims to improve active galactic nuclei (AGN) black hole mass estimates via measurements of the time lag between signatures in continua and emission lines, has fortunately led to increased samples of multi-epoch data. The spectroscopic component of the Australian-based Dark Energy Survey \citep[OzDES and DES respectively;][]{OzDES_description,DES2005,DES2014} began a six year RM program in December 2012 \citep{OzDES_RM_expectations,OzDESFirstDR}. The OzDES RM program is expected to have observed more than 500 quasars ($0.3 < z < 4.5$) at 25 epochs each upon completion. The third and fourth generations of the Sloan Digital Sky Survey \citep[SDSS-III and SDSS-IV;][]{SDSSIIIoverview, Blanton2017} implemented an even larger reverberation mapping program \citep[SDSS~RM;][]{SDSS_RM}. The SDSS~RM program has obtained more than 50 spectra for each of 849 quasars over a similar redshift range to OzDES. 

The SDSS~RM data set is ideal for studying the intrinsic variability of quasars. Multiple observations, a broad span in redshifts, and a broad span in steady-state luminosities enable studies of intrinsic variability as a function of global quasar properties. Additionally, the high redshifts spanned by the RM sample reveals properties of the UV region in quasar accretion disks. This characteristic provides causal insight into variations in emission line strengths and insight into the effects of quasar continuum diversity due to intrinsic variability on cosmology involving quasars' Lyman-$\alpha$ forests.

We use the SDSS RM sample to study spectral variability of quasars. The data are described in Section~\ref{sec:data}. In Section~\ref{sec:diffspec}, we generalize intrinsic variability across our sample of quasars with the creation of a composite differential spectrum. This signature is used to evaluate the logarithmic changes in flux at each wavelength, relative to quasars' steady-state, as a function of the change in the log of luminosity. In Section~\ref{sec:splits}, we produce iterations of the composite differential spectrum using different definitions of luminosity for the quasars, and we discuss the implications of our results on both quasar astrophysics and cosmological studies on the 1D Lyman-$\alpha$ auto-correlation. The work is summarized in Section~\ref{sec:con}.

\begin{table*}[]%[!htb]
        \begin{center}
                \begin{tabular}{c c c c}
                \hline \hline
                Selection Criterion & \# Quasars  & \# Spectra & \% Spectra\\
                 & Remaining & Remaining & Remaining\\
                \hline
                All SDSS RM spectra & 849 & 42,693 & 100\% \\
                Quasar is represented in DR12Q & 835 & 41,993 & 98\% \\
                Quasar in range $1.62 \leq$ {\tt z\_PCA} $\leq 3.30$ & 397 & 19,996 & 47\% \\
                Quasar has no BALs & 342 & 17,232 & 40\% \\
                Spectra have {\tt Z\_WARNING}=0 & 342 & 16,499 & 39\% \\
                $z$ of epochs of the same quasar differ by no more than 0.05 & 342 & 16,433 & 38\% \\
                Spectra have unmasked flux where luminosity derived & 342 & 16,431 & 38\% \\
                Quasar has at least 11 epochs left & 340 & 16,421 & 38\% \\
                \hline
                \end{tabular}
                \caption{The number of quasars and total spectra remaining in our sample after applying each selection criterion to the original SDSS RM sample.}
                \label{table:conditions}
        \end{center}

\end{table*}

\section{\textbf{Data}}\label{sec:data}

This analysis uses data collected as part of the SDSS-III \citep{SDSSIIIoverview} and SDSS-IV \citep{Blanton2017} programs conducted at the Apache Point 2.5-meter Telescope \citep{ApachePoint}. These two programs each feature a major cosmological spectroscopic survey; in the case of SDSS-III, the cosmology component was the Baryon Oscillation Spectroscopic Survey \citep[BOSS;][]{BOSS}.  In SDSS-IV it is the Extended Baryon Oscillation Spectroscopic Survey \citep[eBOSS;][]{eBOSS2016}. BOSS and eBOSS were designed to constrain the cosmological model through observations of galaxies and quasars with the BOSS optical spectrograph \citep{BOSS_Smee} over a wavelength range of 3600~-- 10,400~\AA.

As BOSS neared its final observations in 2014, it became apparent that there was more time available than was required to complete the main cosmological survey. Through a competitive process, the SDSS RM program was one of several programs selected to fill the remaining observation time.

In the SDSS RM project, 849 quasars in the redshift range $0.1<z<4.5$ over 7 deg$^2$ of sky were observed 32 times in the spring of 2014. The quasars were initially identified in several spectroscopic programs and represent a fairly complete sample to $i<21.7$.  The largest source of high-redshift quasars is the BOSS and eBOSS Lyman-$\alpha$ forest cosmology sample, so there is a large overlap between the properties of SDSS RM quasars and the properties of quasars used in SDSS cosmology studies.  These observations were made roughly every four days between January and July.  An additional 20 observations were acquired at a longer cadence in 2015 and 2016 during the first two years of eBOSS. Among other results, this program has already proven its promise through measurements of broad-line H-$\alpha$, H-$\beta$ and Mg~\textsc{ii} lags at $z \ge 0.3$ \citep{Firstlags, GrierRMResults}. All SDSS RM epochs used in our analysis are included in the 14th public data release of SDSS \citep[DR14;][]{DR14}.

\subsection{Sample Selection} \label{subsec:cuts}

After each observation of the SDSS RM field, spectra were extracted from the CCD images and flux calibrated through an automated data processing pipeline. The general process for reduction of spectra in DR14 is described in \citet{pipeline_calibration} and \citet{pipeline_classification}. There were two major revisions to the spectral data processing in DR14 relative to prior versions. The first improvement reduces bias associated with coaddition of many exposures \citep{Hutchinson}; the second addresses the effect of atmospheric differential refraction on quasar spectra calibration \citep{address_ADR_stuff,Treyper}.   Several additional revisions to the data reductions have been made since DR14.  The most important revision leads to a change in the spectral extractions to account for cross-talk between fibers that was found in studies of \textsc{C~iv} broad absorption line variability (Hemler et al. 2018, submitted).  In what follows, we use version v5\_10\_10 of the data processing algorithm to take advantage of this change to the extraction algorithm.

\citet{DRtwelveQ} present a catalog of quasar properties (DR12Q) for the majority of objects observed in the SDSS RM program. We extract {\tt z\_PCA}, {\tt BAL\_FLAG\_VI}, and photometric extinction values from the DR12Q catalog. To guarantee \textsc{C~iv} visibility in a relatively high S/N region of each spectrum, we only use RM objects with a redshift in the range $1.62 \leq$ {\tt z\_PCA} $\leq 3.30$. We then exclude broad absorption line (BAL) quasars as identified by a {\tt BAL\_FLAG\_VI} value equal to one.

The automated classification scheme assigns a {\tt Z\_WARNING} flag to any spectrum where a reliable redshift could not be fit. Spectra with a {\tt Z\_WARNING}$\ne$0 in the DR14 catalog are excluded from our sample. We then match all epochs belonging to the same quasar, but exclude those with a DR14 redshift estimate which deviates from the mean of all epochs by more than $\Delta z = 0.05$. This cut removes spectra with misassigned fiber IDs due to a known problem that appears infrequently during mountain operations.

We next remove spectra with poor flux quality in the region used to determine a monochromatic luminosity (1680~-- 1800~\AA\ in the rest frame, as discussed below). Finally, objects with ten or fewer epochs remaining after these cuts are excluded from the sample. This iterative refinement of our sample is quantified in Table~\ref{table:conditions}. The final sample has 340 quasars and 16,421 total spectra in the redshift range $1.62 \leq z \leq 3.30$.

We correct the final sample of spectra for Galactic extinction according to the model of \citet{Fitzpatrick} and the extinction parameters provided in DR12Q. We adopt the same method as is used in cosmological Lyman-$\alpha$ forest studies \citep[e.g.][]{Delubac_2015, Bautista} to mask pixels known to be contaminated with night air glow produced by de-excitation in \textsc{OH} vibrational modes in Earth's atmosphere. Spectra are truncated blueward of 3800~\AA\ and redward of 9300~\AA\ to increase the average signal-to-noise ratio (S/N) throughout the full wavelength coverage of individual spectra.

Spectra are converted to a fixed rest frame wavelength grid with equal pixel spacing of $10^{-4}$ in $\log\lambda$, consistent with the sampling of the co-added spectra in BOSS and eBOSS. Flux and associated weights are assigned to the nearest pixel to avoid resampling.

\begin{figure*}[]
        \begin{center}
        \includegraphics[width=0.7\textwidth, bb=110 410 530 600]{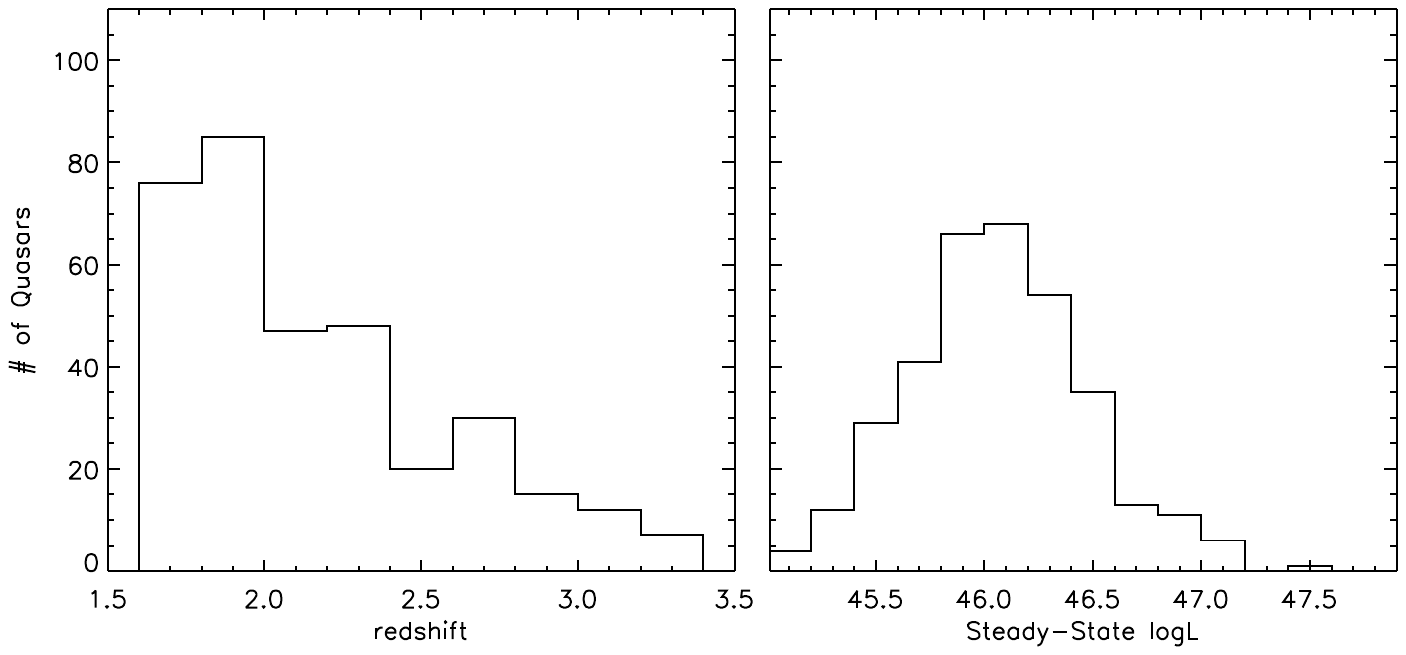} %bb=110 410(405) 530 540
        \includegraphics[width=0.7\textwidth, bb=110 405 530 600]{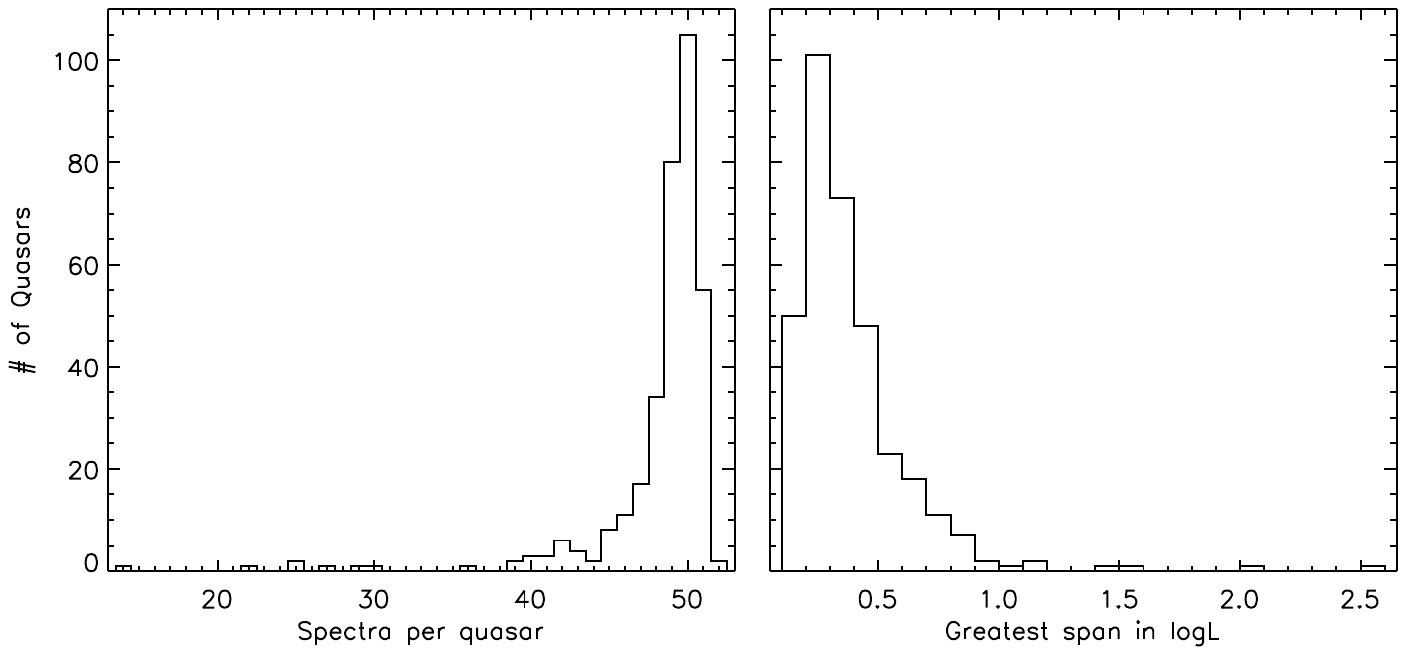} %, bb=100 390 540 575;bb=130 395 520 530
        \caption{Distributions of quasars' redshift, steady-state luminosity,  number of epochs, and greatest span in $\log{L_{\mathrm{bol}}}$ between epochs (i.e., $\log{L_{\mathrm{bol,brightest\_epoch}}} - \log{L_{\mathrm{bol,dimmest\_epoch}}}$).}
        \label{fig:histograms}
        \end{center}
\end{figure*}

\newpage

\subsection{Quasar Properties} \label{subsec:qsoprops} 

Our analysis uses luminosity, spectral index, and \textsc{C~iv} $W_\lambda$ to explore the intrinsic spectral variability of quasars and to assess the impact of quasar variability on Lyman-$\alpha$ forest clustering studies. We compute $L_{\mathrm{bol}}$, $\alpha_{\lambda}$, and \textsc{C~iv} $W_\lambda$ from the spectrum of every epoch of every quasar. The median $\log{L_{\mathrm{bol}}}$, $\alpha_\lambda$, and \textsc{C~iv} $W_\lambda$ for each quasar are considered to represent the quasar's steady state properties.

The bolometric luminosity of each spectrum is estimated from the monochromatic luminosity at 1740~\AA\ (i.e., the median flux in the range 1680~-- 1800~\AA) corrected for luminosity distance under a $\Lambda \mathrm{CDM}$ cosmology with $H_0 = 70\ \mathrm{km \, s^{-1} \, Mpc}$, $\Omega_{M} = 0.3$, and $\Omega_{\Lambda} = 0.7$. We approximate $L_{\mathrm{bol}}=A*L_{1740}$, where we compute the bolometric luminosity at 1450 \AA\ with the correction factor suggested by \citet{Lbol_correction} and determine the scaling factor $A=4.28$ that minimizes the residual between the bolometric luminosities computed at 1740~\AA\ and 1450~\AA.  We note that the analysis that follows uses the fractional change in luminosity, so the choice of scaling of monochromatic flux to bolometric luminosity does not impact any of the results.

The spectral index of each spectrum is determined by fitting a power law to the quasar continuum in the wavelength ranges 1680~-- 1800~\AA\ and 2000~-- 2050~\AA. These ranges were chosen for their lack of emission lines (except broadband iron which we assume is negligible) according to a high S/N composite spectrum \citep{Harris2016}. Every spectrum in our sample is observed over this wavelength range. Only unmasked pixels with flux within three standard deviations of the median are used for this fit to mitigate the influence of intervening absorption from the intergalactic medium.

To calculate \textsc{C~iv} $W_\lambda$ for each spectrum, we first estimate the continuum around \textsc{C~iv} emission with a linear fit to the spectrum in the wavelength ranges 1450~-- 1465~\AA\ and 1685~-- 1700~\AA. We model the \textsc{C~iv} emission using a double Gaussian fit to the continuum-subtracted spectrum over the wavelength range 1500~-- 1580~\AA. Only unmasked flux measurements within three standard deviations of the best fit are used. \textsc{C~iv} $W_\lambda$ is finally computed by taking the integral of the estimated emission flux relative to the estimated continuum flux over 1520~-- 1580~\AA.

The distributions of quasar properties are presented in Figure~1. The four panels represent the redshift distribution, steady-state luminosities, epochs per quasar, and span between minimum and maximum $\log{L_{\mathrm{bol}}}$.

\subsection{Linear Trends in Quasar Properties}
\label{Sec:scatterplots}
In this subsection, we explore the Baldwin and bluer-means-brighter effects in our data. We fit linear relations to these data, but do not provide errors on measured slopes because the measurements are dominated by intrinsic scatter. Instead, we report the rms scatter about the fit.

The top panel of Figure~2 shows the global Baldwin Effect for \textsc{C~iv} for the 340 quasars in the sample. The relationship between equivalent width and luminosity is shown for every epoch, where individual quasars have a fixed color scheme. The median $\log{(\textsc{C~iv} \, W_\lambda)}$ and $\log{L_{\mathrm{bol}}}$ of each quasar are shown in black, the collection of which represents the global BE with minimal scatter from intrinsic variation. We obtain a fit to the linear relationship between these steady-state $\log{(\textsc{C~iv} \, W_\lambda)}$ and $\log{L_{\mathrm{bol}}}$ values that yields a slope of $-0.272$ for the global BE. This observed relation is steeper than those observed in \citet{Kinney} and \citet{Wilhitecivline}, and shallower than that observed in \citet{Treyper}. As discussed in \citet{Treyper}, the differences likely depend on sample properties such as redshift.

Targeting only the intrinsic BE, the bottom panel of Figure~2 shows the deviation in $\log{(\textsc{C~iv} \, W_\lambda)}$ and $\log{L_{\mathrm{bol}}}$ of each quasar spectrum relative to the steady state. Iterative clipping was used in a linear regression fit to define the slope for the intrinsic BE. Data more than three standard deviations from the best fit were rejected until the fits converged to within 0.002. The final linear fit has a slope of $-0.695$. Data which were excluded from the fit are marked with `x's, and the color scheme is consistent with that used for the global BE. The slope for the intrinsic BE is roughly 2.5$\times$ that of the global BE. The steeper relationship for intrinsic variation is consistent with previous findings \citep{Kinney,Wilhitecivline}.

We also conduct linear analyses of the global and intrinsic relationships of spectral index vs. luminosity (Figure~3). The color scheme and fitting techniques in the top two panels are the same as for Figure~2. These correlations provide alternative but consistent perspectives on the intrinsic bluer-means-brighter trend. The slope for the global relationship between $\alpha_\lambda$ and $\log{L_{\mathrm{bol}}}$ (top panel) is nearly zero (0.017) across a luminosity range spanning more than two dex. This result is consistent with no apparent dependence of color on steady-state luminosity \citep{photometric_analog}. The data in the middle panel of Figure~3 gives a slope of $-1.673$, indicating a transition to bluer (redder) spectra as quasars increase (decrease) in luminosity. This trend is similar to that seen in previous works \citep{Wilhite_wavelength, Pereyra2006, photometric_analog}.

\begin{figure}[tb]
        \begin{center}
        \includegraphics[width=0.49\textwidth]{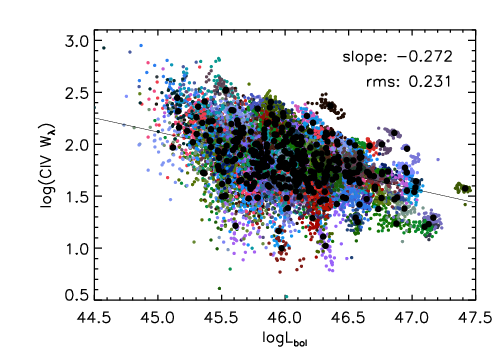} %, bb=75 365 550 725
        \includegraphics[width=0.49\textwidth]{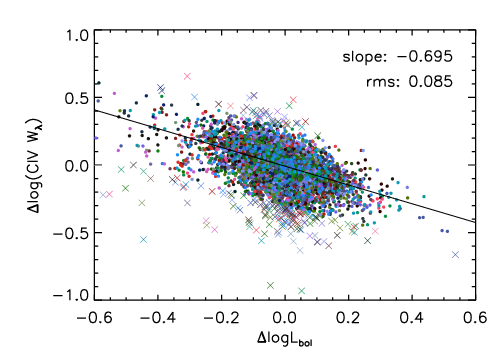} %, bb=75 365 550 725
        \caption{\textbf{Top:} $\log{(\textsc{C~iv}~W_\lambda)}$ vs. $\log{L_{\mathrm{bol}}}$ for every epoch of every quasar. Epochs belonging to the same quasar have the same color. The steady-state $\log{(\textsc{C~iv}~W_\lambda)}$ and $\log{L_{\mathrm{bol}}}$ of each quasar is overplotted in black. A line is fit to these steady-state data, targeting the global Baldwin Effect with minimal added scatter from intrinsic variability. Its slope and root mean square (rms) are shown. \textbf{Bottom:} Targeting only the intrinsic BE, the bottom panel shows the differences in epochs' $\log{(\textsc{C~iv}~W_\lambda})$ from the steady-state $\log{(\textsc{C~iv}~W_\lambda)}$ of the quasars to which they belong, vs. the same differences in $\log{L_{\mathrm{bol}}}$. As described in the text, a line is fit to the data after outlier rejection (outliers marked with `x's) and represents the intrinsic Baldwin Effect. Its slope and rms are shown in the top corner.}
        \label{fig:EWL}
        \end{center}
\end{figure}

\begin{figure}[tb]
        \begin{center}
        \includegraphics[width=0.49\textwidth]{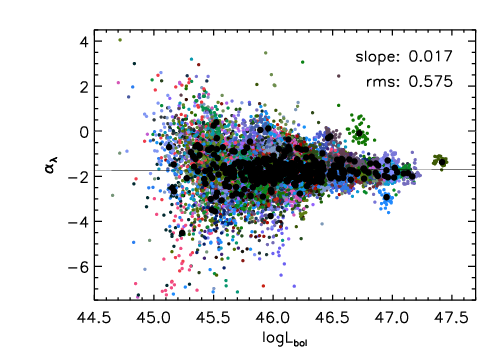} %, bb=75 365 550 725
        \includegraphics[width=0.49\textwidth]{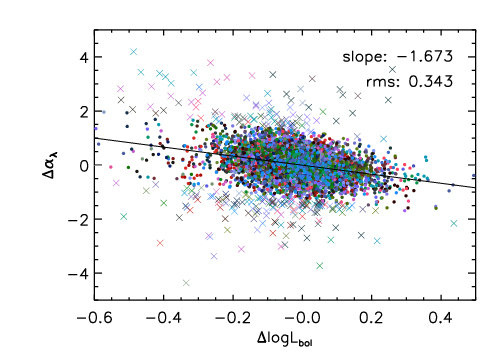} %, bb=75 365 550 725
        \includegraphics[width=0.49\textwidth]{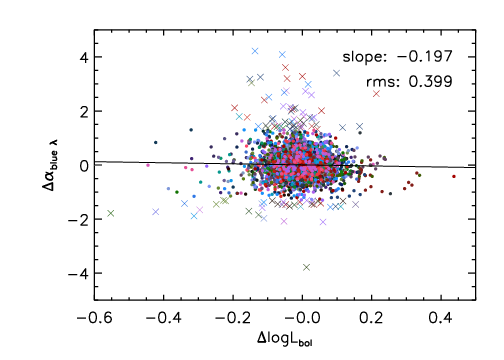}
        \caption{\textbf{Top and Middle:} Spectral index vs. $\log{L_{\mathrm{bol}}}$ and the change in spectral index vs. the change in $\log{L_{\mathrm{bol}}}$ for all epochs of all quasars. Formatting and line fitting in these plots are the same as in the top and bottom panels of Figure~2, respectively. \textbf{Bottom:} Same as the middle panel, but corresponding to the continuum at shorter wavelengths as referenced in Section~3.1. $\alpha_{\mathrm{blue} \lambda}$ is calculated from flux in the range 1130~-- 1165~\AA\ and 1440~-- 1480~\AA. 92 quasars ($z \geq 2.36$) are observed with this wavelength coverage and are shown. Data are color-coordinated by quasar, though quasars' colors are not necessarily the same as they are in previous plots. A line is fit to the data in the same manner as in the middle panel.   \\}
        \label{fig:AL}
        \end{center}
\end{figure}

\newpage

\section{\textbf{The Composite Differential Spectrum}}\label{sec:diffspec}

\begin{figure}[!b]
        \begin{center}
                \includegraphics[bb=80 360 575 705, width=0.5\textwidth]{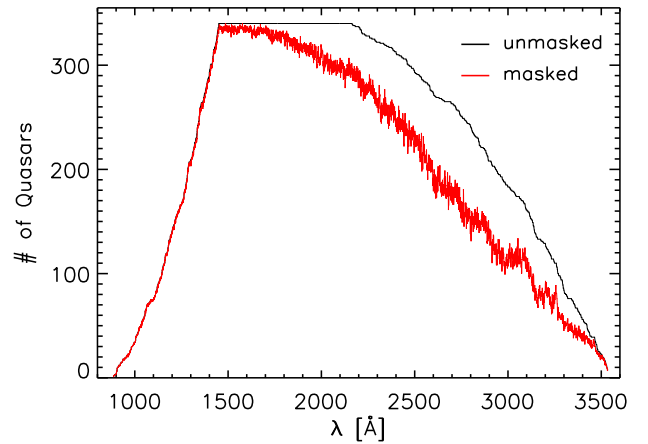}
                \caption{The number of quasars contributing to the composite differential spectrum in Figure~\ref{fig:logdfdl} at each pixel in the restframe wavelength grid (red). Also shown is the number of pixels that would have contributed to the composite differential spectrum if no pixels had been masked (black).}
                \label{fig:pixels}
        \end{center}
\end{figure}

Differential spectra offer more detailed insight into color variation and couplings between the continuum and emission line properties than do the linear relationships explored in Section~\ref{Sec:scatterplots}. We determine the spectroscopic signature of intrinsic variability relative to the steady state spectrum ($F(\lambda)_{\mathrm{steady-state}}$) and steady state luminosity ($L_{\mathrm{bol,steady-state}}$) by computing the quantity
\begin{equation}
\frac{\Delta \log{F}}{\Delta \log{L_{\mathrm{bol}}}}(\lambda)=\frac{\log{F(\lambda)}-\log{F(\lambda)_{\mathrm{steady-state}}}}{\log{L_{\mathrm{bol}}}-\log{L_{\mathrm{bol,steady-state}}}}.
\end{equation}
We constrain the term $m(\lambda) = \frac{\Delta \log{F}}{\Delta \log{L_{\mathrm{bol}}}}(\lambda)$ for each of the 340 individual quasars using the spectrum ($F(\lambda)$) and luminosity ($L_{\mathrm{bol}}$) from each epoch.  As the black hole mass is fixed across epochs of the same quasar, this signature captures spectral responses to changing mass accretion rates, or $R_{\mathrm{Edd}}$, at a fixed black hole mass.

Directly computing the quantity using $\log{F(\lambda)}$ found in Equation~1 is complicated by low signal-to-noise measurements.  Instead, we rely on the linear flux measurements and associated errors to determine the spectral signature of variability.  For each quasar, at every pixel in the quasar rest frame, we fit the following relationship
\begin{equation}
F(\lambda)_i=F(\lambda)_{\mathrm{steady-state}}*10^{m(\lambda) \Delta \log{L}_{\mathrm{bol},i}}.
\end{equation}
$F(\lambda)_i$ is the observed flux at each epoch $i$ and is the dependent variable. $\Delta \log{L}_{\mathrm{bol},i}$ is the independent variable and the same quantity that was plotted in Figures~2 and 3. $m(\lambda)$ and $F(\lambda)_{\mathrm{steady-state}}$ are free parameters; $m(\lambda)$ is equal to $\frac{\Delta \log{F}}{\Delta \log{L_{\mathrm{bol}}}}(\lambda)$ and thus represents the desired signature at each pixel. $F(\lambda)_{\mathrm{steady-state}}$ represents the average flux of that quasar over the duration of the SDSS RM program. The associated measurement errors for $F(\lambda)_i$ are provided in v5\_10\_10 of the eBOSS data and were used for this fit.

\begin{figure*}[!t]
        \begin{center}
                \includegraphics[bb=130 355 510 715]{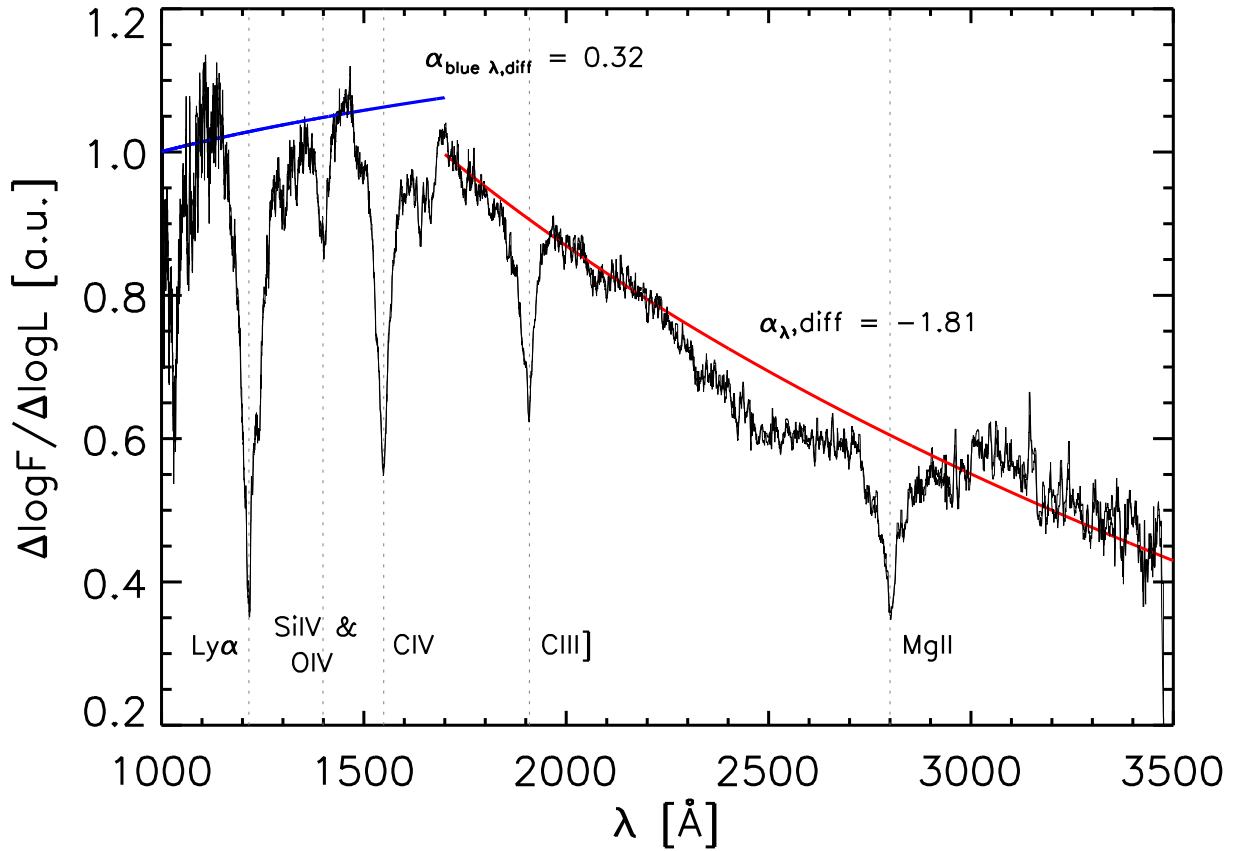} %width=0.85\textwidth,, bb=75 365 550 725
                \caption{The median-combined composite differential spectrum made from all 340 quasars in arbitrary units, smoothed with a boxcar of 10 pixels (black). A power law with slope $-1.81$ (red) is fit in the regions used to calculate each spectrum's spectral index, i.e. 1680~-- 1800~\AA\ and 2000~-- 2050~\AA. The power law fit to the regions 1130~-- 1165~\AA\ and 1440~-- 1480~\AA\ (blue) has slope $\alpha_{\mathrm{blue} \lambda,\mathrm{diff}}=0.32$.}
                \label{fig:logdfdl}
        \end{center}
\end{figure*}

To generalize the signature of variability across the quasar population, we co-add the unmasked $\frac{\Delta \log{F}}{\Delta \log{L_{\mathrm{bol}}}}(\lambda)$ terms from each individual quasar into a median-combined differential spectrum. As intrinsic variations in the quasar sample dominate the uncertainties, we do not report the statistical measurement errors on this spectrum.  Figure~\ref{fig:pixels} shows the number of quasars contributing to the composite differential spectrum at every pixel. The composite differential spectrum, smoothed with a boxcar average of 10 pixels, is shown in Figure~5. This result generalizes intrinsic variability with respect to luminosity across our full sample and is the main contribution of this work.

Figure~\ref{fig:logdfdl} shows a suppression in $\frac{\Delta \log{F}}{\Delta \log{L_{\mathrm{bol}}}}$ for all prevalent emission lines relative to the continuum. This signature is qualitatively in line with the intrinsic BE, as it implies that emission line flux increases with luminosity to a lesser degree than continuum flux (i.e. the equivalent width of emission lines decreases with increasing luminosity).

\begin{table*}%[!t] color coding
        \begin{center}
                \caption{
                \label{table:bins}
                Characterization of the BE and color trends as a function of wavelength range used to determine the change in luminosity.}
                \tablenotetext{a}{The wavelength region $1375 < \lambda < 1425$ \AA\ is excluded to avoid contamination from the \textsc{O~iv}/\textsc{Si~iv} emission lines.}
                \begin{tabular}{c c c c c c c}
                \hline \hline
                Wavelength Range & $\lambda_{\rm eff,Lum}$ & Global BE & Intrinsic BE & Differential Power Law & Differential Power Law & Fractional Dispersion \\
                (\AA) & (\AA) & Slope & Slope & Slope ($\lambda >1680$ \AA) & Slope ($\lambda <1480$ \AA) & in Continuum Flux \\
                \hline
                1050--1150 & 1100 & -0.28 & -0.34 & -0.72 & -2.11 & 0.172 \\
                1300--1500\tablenotemark{a} & 1400 & -0.25 & -0.65 & -2.13 & -1.56 & 0.101 \\
                1680--1800 & 1740 & -0.27 & -0.70 & -1.81 & 0.33 & 0.108 \\
                2000--2200 & 2100 & -0.27 & -0.67 & 0.05 & 1.19 & 0.100 \\
                1400--2500 & $g+r$& -0.26 & -0.75 & -1.40 & 0.28 & 0.057 \\
                \hline
                \end{tabular}
        \end{center}
\end{table*} 

\subsection{Color Variation in the Differential Spectrum}

The composite differential spectrum shows continuum structure consistent with the bluer-means-brighter effect: the continuum strength around 2000~\AA\ generally increases more than the continuum strength around 3000~\AA. Arising from the Rayleigh Jeans tail of observed quasar spectra, a power law (with slope $\alpha_{\lambda, \mathrm{diff}}$; hereforth referred to as ``the differential power law'' ) can be used to characterize this structure.  Fitting to the differential continuum in Figure~\ref{fig:logdfdl} over the same regions used to find each epoch's spectral index (1680~-- 1800~\AA\ and 2000~-- 2050~\AA), we obtain a differential power law of slope $\alpha_{\lambda, \mathrm{diff}}=-1.81$. This fit is illustrated in red in Figure~\ref{fig:logdfdl}, and is similar to those of the full residual spectra (1300~-- 6000~\AA) in \citet{Pereyra2006} and \citet{Ruan2014}.
$\alpha_{\lambda, \mathrm{diff}}$ represents the degree to which the spectral index of a single object changes with an order of magnitude change in luminosity, making it qualitatively identical to the measurement of the slope in the middle panel of Figure~\ref{fig:AL}; the slope $\alpha_{\lambda, \mathrm{diff}}=-1.81$ is reasonably consistent with the slope of $-1.67$.

The surprising feature in our differential spectrum is an apparent break in the power law blueward of 1700~\AA. We verify this break by recreating the linear $\Delta\alpha_{\lambda}$ vs. $\Delta \log{L_{\mathrm{bol}}}$ relationship using data blueward of 1700~\AA\ to perform the spectral index fit for every epoch. The results are shown in the bottom panel of Figure~\ref{fig:AL}. In this representation, $\alpha_{\mathrm{blue} \lambda}$ is the spectral index of each spectrum as calculated from the regions 1130~-- 1165~\AA\ and 1440~-- 1480~\AA. 92 quasar observations ($z \geq 2.36$) have coverage in this wavelength range. A linear fit is made to the trend of $\Delta\alpha_{\mathrm{blue} \lambda}$ versus $\Delta \log{L_{\mathrm{bol}}}$ with the same iterative process described in Section~\ref{Sec:scatterplots}, leading to a best fit slope equal to $-0.197$. We compare this slope to the corresponding ``power law'' in the differential spectrum measured over the same wavelength range. From the differential spectrum, we find $\alpha_{\mathrm{blue} \lambda,\mathrm{diff}}=0.32$ as shown in blue in Figure~\ref{fig:logdfdl}. These two measurements are consistent in that both reflect a break from the steeper power law found at wavelengths longer than 1700~\AA.  Discrepancies in the absolute slopes likely arise from degradation in the quality of the fit to $\alpha_{\mathrm{blue} \lambda}$ in individual quasar spectra due to the Lyman-$\alpha$ forest.

\section{\textbf{Interpretation}}\label{sec:splits}

In this section we explore the astrophysical origins and cosmological implications for the intrinsic variability of quasars presented in Section~\ref{sec:diffspec}. We begin by assessing the signature of variability across different rest frame wavelengths used to estimate the change in bolometric luminosity.  Next, we qualitatively discuss our results in the context of existing quasar accretion disk models. Finally, we quantify the improvements that our differential spectrum (Figure~\ref{fig:logdfdl}) can pose for cosmology analyses that use quasar spectra's Lyman-$\alpha$ forests to probe density fluctuations of neutral hydrogen in the Universe.

\subsection{Dependence of Differential Spectrum on Luminosity Definition}

The interpretation of the continuum shape in the differential spectrum shown in Figure~\ref{fig:logdfdl} depends on how reliably
the bandpass of $1680 < \lambda < 1800$ \AA\ predicts the flux across the full spectral range covered by the BOSS spectrographs.
If the trend and break in the power law are universal, then a model for quasar accretion must account for the suppression of
flux at shorter wavelengths with increasing accretion rate.
One can then incorporate the differential spectrum as a template in the modeling of the continuum level in the Lyman-$\alpha$
forest using fits to the unabsorbed spectrum at rest frame wavelengths greater than 1216 \AA.
On the other hand, the location of a break just blueward of the wavelengths used to define luminosity hints at a
coincidence that requires further investigation.
Here, we compute composite differential spectra using other definitions of luminosity.  We use these composite spectra
to test whether the signature in Figure~\ref{fig:logdfdl} is a comprehensive description of quasar variability,
or instead, if the observed signature is particular to the choice of bandpass used to assess the change in luminosity.

We start by identifying a series of bandpasses that are relatively free of emission lines based on visual inspection of the
composite spectrum presented in \citet{Harris2016}. 
We use the wavelength ranges found in the left column of Table~\ref{table:bins}, thus covering more than a factor of two in restframe
wavelength.
We reject the region 1375--1425 \AA\ because the luminosity estimate in bin $1300 < \lambda < 1500$ \AA\ would otherwise
be contaminated by flux from the \textsc{O~iv}/\textsc{Si~iv} emission lines.
The largest bin covering $1400 < \lambda < 2500$ \AA\ does not exclude any regions and is intended to approximate
a very broad imaging filter (e.g., $g+r$ at a redshift $z=1.9$).
For each wavelength range, we define the luminosity for each quasar at each epoch exactly as outlined in Section~2.2.
We introduce the term $\lambda_{\rm eff,Lum}$ to denote the effective wavelength used to estimate the luminosity for each bin.
We preserve the definitions of spectral index and \textsc{C~iv} $W_\lambda$ from Section~2.2.

When assessing the trend of \textsc{C~iv} $W_\lambda$ with the new definitions of luminosity, we are able to reproduce the global BE slopes
found in the top panel of Figure~\ref{fig:EWL}.  The best fitting slope of the global \textsc{C~iv} $W_\lambda$ versus $\log{L_{\mathrm{bol}}}$ relationship
for each bin is reported in the third column of Table~\ref{table:bins}.  The slope varies from the original result by at most
two parts in 27, thus implying that the population of quasars in this study follows a global Baldwin relationship that does not strongly depend on the definition of luminosity.
However, when assessing the intrinsic BE, as in the bottom panel of Figure~\ref{fig:EWL}, the slope of the
relationship changes significantly.  As shown in the fourth column of Table~\ref{table:bins}, the equivalent width of \textsc{C~iv}
follows a much shallower trend at $\lambda_{\rm eff,Lum}=1100$ \AA\ than when a bandpass at longer wavelengths is used to determine luminosity.

\begin{figure*}[!t]
        \begin{center}
        \includegraphics[width=0.9\textwidth, bb=125 375 520 700]{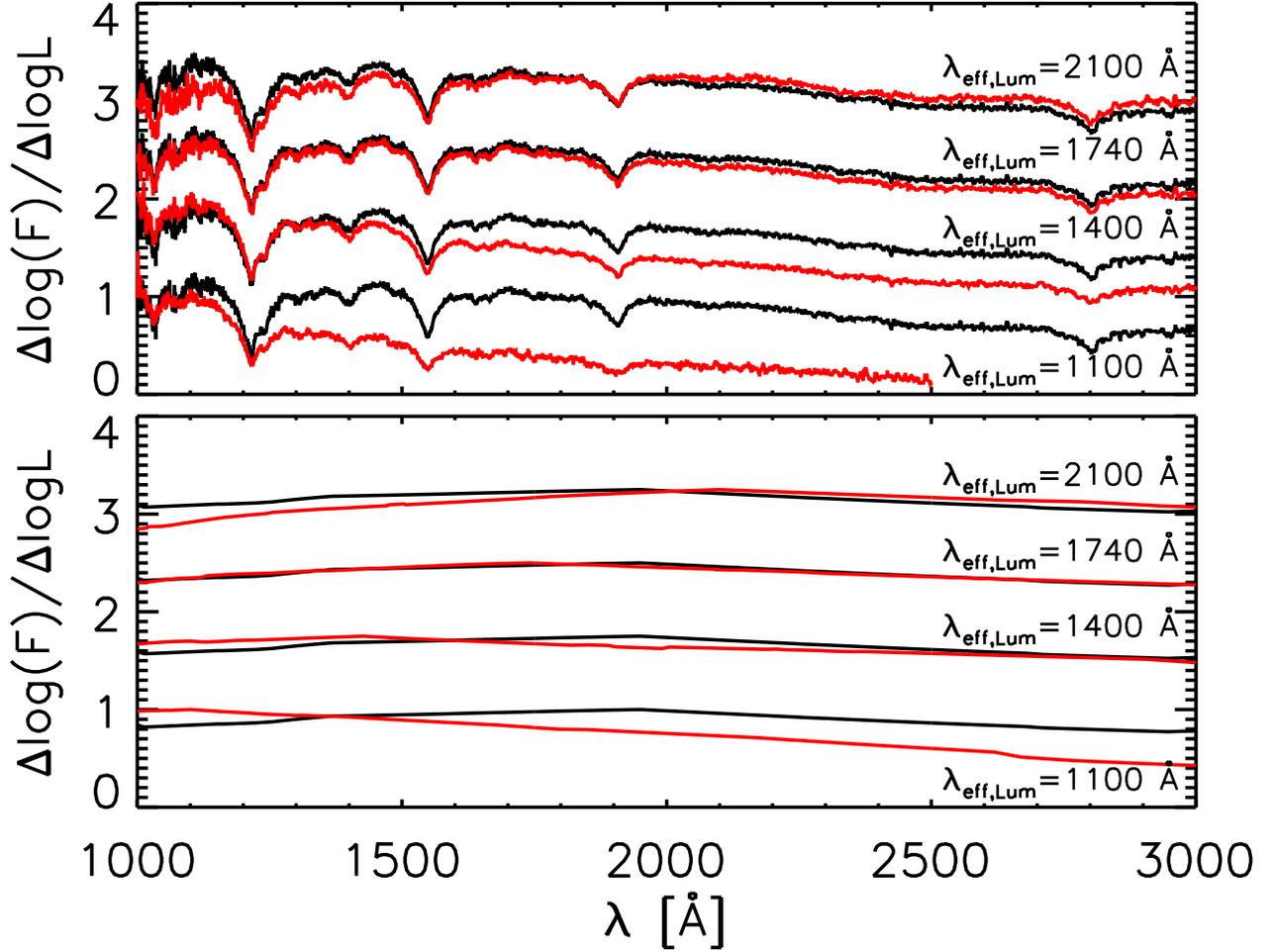} %, bb=125 385 520 700, bb=125 390 515 710, , bb=68 25 472 340
\hspace{0.5cm}
        \caption{{\bf Top:  }Observed composite differential spectra computed with different definitions of luminosity.
All curves are smoothed with a boxcar width of 10 pixels.
{\bf Bottom:}  Differential spectra simulated according to the inhomogeneous accretion disk model using different definitions of luminosity.
In both panels, the black curve represents the $\lambda_{\rm eff,Lum}=g+r$ differential spectrum.
Each black curve is identical and is offset by 0.75 dex to provide a reference for the other curves.
Each red curve represents the differential spectrum with luminosity computed over a more narrow wavelength interval, offset by 0.5 dex for clarity.
From top to bottom, the curves represent the differential spectra computed with the $\lambda_{\rm eff,Lum}=2100$ \AA, $\lambda_{\rm eff,Lum}=1740$ \AA,
$\lambda_{\rm eff,Lum}=1400$ \AA, and $\lambda_{\rm eff,Lum}=1100$ \AA\ definitions of luminosity.
In the observed data, the $\lambda_{\rm eff,Lum}=1400$ \AA\ bin excludes all flux with $1375 < \lambda < 1425$ \AA\ in order to avoid the \textsc{O~iv}/\textsc{Si~iv} emission lines.
}
        \label{fig:binned_dfdl}
        \end{center}
\end{figure*}

We then produce a composite differential spectrum for each luminosity bin following the framework presented in Section~3.
The results are shown in the top panel of Figure~\ref{fig:binned_dfdl}.
The differential spectrum computed under the $\lambda_{\rm eff,Lum}=g+r$ assumption is presented in black to provide a reference for
every other composite spectrum.  The composite spectra for the other bins are presented from top to bottom in order of luminosity defined by decreasing wavelength.
The shape of the continuum varies significantly depending on the choice of bandpass used for estimating luminosity. 
Namely, when using $\lambda_{\rm eff,Lum}=2100$ \AA, one sees an overall flattening of the spectrum relative to the broad bandpass estimate.  When using $\lambda_{\rm eff,Lum}=1100$ \AA,
one sees a much steeper spectrum at wavelengths $\lambda < 1300~$\AA\ and a deficit of signal at longer wavelengths relative to the broad bandpass estimate.
In general, the differential spectra peak around the wavelength range used to determine the luminosity, with the possible exception of the bin 
$\lambda_{\rm eff,Lum}=1400$ \AA.  

To quantify these trends, we determine power law fits to the same regions of each spectrum as those shown
in Figure~\ref{fig:logdfdl}.  The slopes of the power laws for each composite spectrum over the wavelengths 1680~-- 1800~\AA\ and 2000~-- 2050~\AA\ are presented
in the fifth column of Table~\ref{table:bins}.  The slopes of the power laws over the wavelengths 1130~-- 1165~\AA\ and 1440~-- 1480~\AA\ are presented in
the sixth column.  Here, the difference between each differential composite spectrum becomes most clear.  The differential spectra grow
bluer (more negative slope) the closer the fitting range is to the wavelength range used to compute luminosity.  Each bin argues for a break in the power law,
but the location and the strength of the break is strongly influenced by the choice of the wavelength range used to determine luminosity.

The differential spectra demonstrate that the signature of variability depends on which wavelength range is used to characterize the quasar brightness.
The wavelengths that are proximate to the wavelength region used to measure luminosity experience the largest change in flux with the change in luminosity.
This effect leads to an apparent peak at these locations and a break in the power law at wavelengths further from the luminosity window.
The variation in signature implies that any one region of the quasar spectrum cannot fully predict the features across the full range
of $1000 < \lambda < 3000 $ \AA.  We compute the dispersion and correlation coefficients for integrated flux within each bandpass.
The dispersion of all flux measurements relative to the steady-state flux of each quasar for each wavelength bin is shown in the last column of Table~\ref{table:bins}.
One notes that the dispersion in the continuum in the Lyman-$\alpha$ forest region is much larger than in the other bins.
The amplitude of the differential spectrum over the Lyman-$\alpha$ forest region is comparable to or less than the 
amplitude at redder wavelengths for the $\lambda_{\rm eff,Lum}=1740$ \AA\ and $\lambda_{\rm eff,Lum}=2100$ \AA\ bins. These two templates 
therefore underestimate the variability in the quasar spectra at the shortest wavelengths.  One also notes that the dispersion of the flux is 
lowest for the $1400 < \lambda < 2500$ \AA\ bin. The amplitude of the variability of this wide bin must be suppressed due to the broad wavelength coverage.

The observation that the variations in quasar flux become decorrelated with increasing separation in wavelength
is further supported by the following correlation matrix:

\begin{equation}
Corr(i,j) =
\begin{pmatrix}
 1 & 0.814 & 0.669 & 0.478 \\
   & 1     & 0.891 & 0.716 \\
   &       & 1     & 0.844 \\
   &       &       &  1    
\end{pmatrix} 
.
\end{equation}
The elements in the correlation matrix are ordered according to the flux measured over the intervals
$1050 < \lambda < 1150$ \AA, $1300 < \lambda < 1500$ \AA, $1680 < \lambda < 1800$ \AA, and $2000 < \lambda < 2200$ \AA.
In all cases, the correlation coefficient decreases at increasing separation in wavelength.
The processes that lead to variability in quasar luminosity produce a dominant component of flux variability that is local to a narrow range of wavelengths.

\subsection{Quasar Accretion Disk Models}

It is possible to attribute the observed continuum behavior in the top panel of Figure~\ref{fig:binned_dfdl} and in
Equation~3 to several accretion disk models. For example, the increase in luminosity may be enhanced over narrow regions of wavelength as prescribed
by the inhomogeneous disk model. The signature could also result from reverberation in the accretion disk during a transition of disk structure towards
a slim disk as objects increase in luminosity, or equivalently Eddington efficiency. Both of these scenarios also predict the intrinsic Baldwin Effect as seen in our differential spectrum and Figure~\ref{fig:EWL}.

{\bf Inhomogeneous accretion disk:}
Under the inhomogeneous disk model \citep{DexterAgol}, the observed signature of variability is a superposition of ephemeral
blackbody emission from independently-varying zones.
The spectrum at each epoch depends on the two free parameters of $n$ and $\sigma_{T}$, where $n$ is the number of localized zones in the accretion disk per factor of two in radius
and $\sigma_{T}$ is the amplitude of their temperature fluctuations in log space. As shown in Figures~6 and 7 of \citet{Ruan2014}, the relative variability spectrum can produce a
break in the power law similar to that shown in Figure~\ref{fig:logdfdl} when certain combinations of $n$ and $\sigma_{T}$ are used.  Generally, the blackbody peak from a given zone occurs
farther into the UV with increasing $\sigma_{T}$, thus pushing the break in the differential spectrum to shorter wavelengths.  The differential power law turns over at redder wavelengths
for smaller $n$ because flux variability will be dominated by flares in the lower-temperature regions of the accretion disk.

The consistency between the simulated variability spectra from \citet{Ruan2014} and our differential spectrum provides compelling evidence for stochastic variation in temperature under the inhomogeneous model.
The model predicts temperatures that vary in an uncorrelated fashion at different radii, thus leading to fluctuations in the peak blackbody
curve at a given radius.  The resulting fluctuation in emission from that region is maximally correlated at wavelengths near the peak of the blackbody curve;
the correlation decreases at increasing separation in wavelength. This mechanism could explain the break in the differential power law and why the position of the break depends on which wavelength region is used
to characterize the variability in bolometric luminosity.

We reproduce simulated differential spectra from the inhomogeneous disk model to further test the change in spectroscopic
signature with definition of luminosity.
We set up the inhomogeneous accretion disk model following the procedure of \citet{Ruan2014}. Assuming a face
on disk, we divided the disk into several zones which are
equal in size radially and azimuthally. Radially, the disk
extends to $10^{20}$ cm with each zone spanning
0.2 dex width. In the azimuthal direction, we divide the disk
into eight equal slices. We assumed a black hole mass of $10^9$ $M_\odot$ and mass accretion rate, $\dot{M}_\mathrm{BH}$,
of 2 $M_\odot$/year. The logarithmic temperature (log(T)) in each zone
independently fluctuates as a first-order damped random
walk process. The mean temperature in each zone is set to
the log(T) of the standard thin accretion disk model at that
radius and a value of 0.6 is assumed for $\sigma_T$. The characteristic
decay timescale, $\tau$, of the temperature fluctuations is set to
200 days.  After a burn-in time of 500 days, we
construct the spectrum by summing the intensity from each
zone according to its blackbody curve and area.
We repeat this procedure 100 times to create 100 realizations
of spectra for individual quasar epochs. In these 100 realizations, we
hold the ${M}_\mathrm{BH}$, $\dot{M}_\mathrm{BH}$, $\sigma_T$ and $\tau$ parameters constant. For each
of the 100 spectra, we record the intensity over the broad wavelength range $1400<\lambda<2500$ \AA, and over the same
narrow ranges centered at 1100 \AA, 1400 \AA, 1740 \AA, and 2100 \AA\ as those in Table~\ref{table:bins}.
We use these intensities to determine differential spectra as was done in Section~4.1.

The differential spectra produced from the simulated quasar spectra are shown in the bottom panel of Figure~\ref{fig:binned_dfdl}.
The results are arranged in the same manner as in the top panel for comparison to the observed differential spectra.
All of the trends identified in the data are reproduced in these simulations, albeit with differing amplitudes.  The most
prominent trend is the migration of the break in the power law toward shorter wavelengths when luminosity is determined
at shorter wavelengths.  The commonality between the features in the simulated spectra
and the features in the observed data suggests that inhomogeneities in the accretion disk
are responsible for the signature of variability observed in the SDSS RM data.
While it may be possible to better reproduce the full shapes of the observed differential spectra in all bins by tuning all five parameters, the parameter
space of the simulated data is degenerate and not entirely physical.  We therefore refrain from further tuning of the model.

{\bf Slim accretion disk:} The signature seen in the top panel of Figure~\ref{fig:binned_dfdl} may alternatively
reflect slim accretion disk structure \citep{SlimDisk_old} with possible truncation of the inner accretion disk
due to line-driven winds \citep{Jiang_simulations, LaorAndDavis}. In the slim disk model, an optically-thick
shielding gas partially obstructs X-ray and far UV (FUV) light radiating from the disk
(see Figure~15 in \citet{leighly04} and Figure~18 in \citet{LuoBrandt2015}).
This suppression in high energy continuum flux reduces ionization and resonant emission in the broad line region,
and suppresses X-ray/FUV continuum strength in observed spectra if viewed with a high inclination angle.
These suppressions will be most prevalent for near-Eddington accretion disks, but likely appear across a
continuum of $R_{\mathrm{Edd}}$ values, strengthening with Eddington efficiency.
\citet{LuoBrandt2015} present evidence of slim disk structure in an extreme sample of high Eddington X-ray weak objects.
Figure~10 in that work shows what appears to be a power law break around 1700~\AA\ for single-observation X-ray weak quasars,
though this characteristic is not specifically discussed.

Even though we do not report accretion rates for this sample, our differential
signature does trace spectral responses to fluctuating Eddington efficiencies.
Any perturbations in the covering factor of shielding gas under the slim disk paradigm would appear in our spectra
as the intrinsic Baldwin Effect. Additionally, if the temperature of the shielding gas decreases as it thickens or
expands to larger radii, or if the increased accretion rate is associated with strengthening winds that truncate
the inner accretion disk, then one would also expect a break in the differential power law.
Finally, it is possible that the decorrelation of intensity fluctuations with wavelength separation is a result of delays in the
response of the outer blackbody disc to changes in illumination from the inner disk \citep[as in ][]{gardner17a}.

\subsection{Implications for Cosmology using Lyman-$\alpha$ Forests}\label{subsec:cosmology}

\begin{figure*}[tb] %cosmology
        \begin{center}
                \includegraphics[bb=85 355 570 715, width=0.49\textwidth]{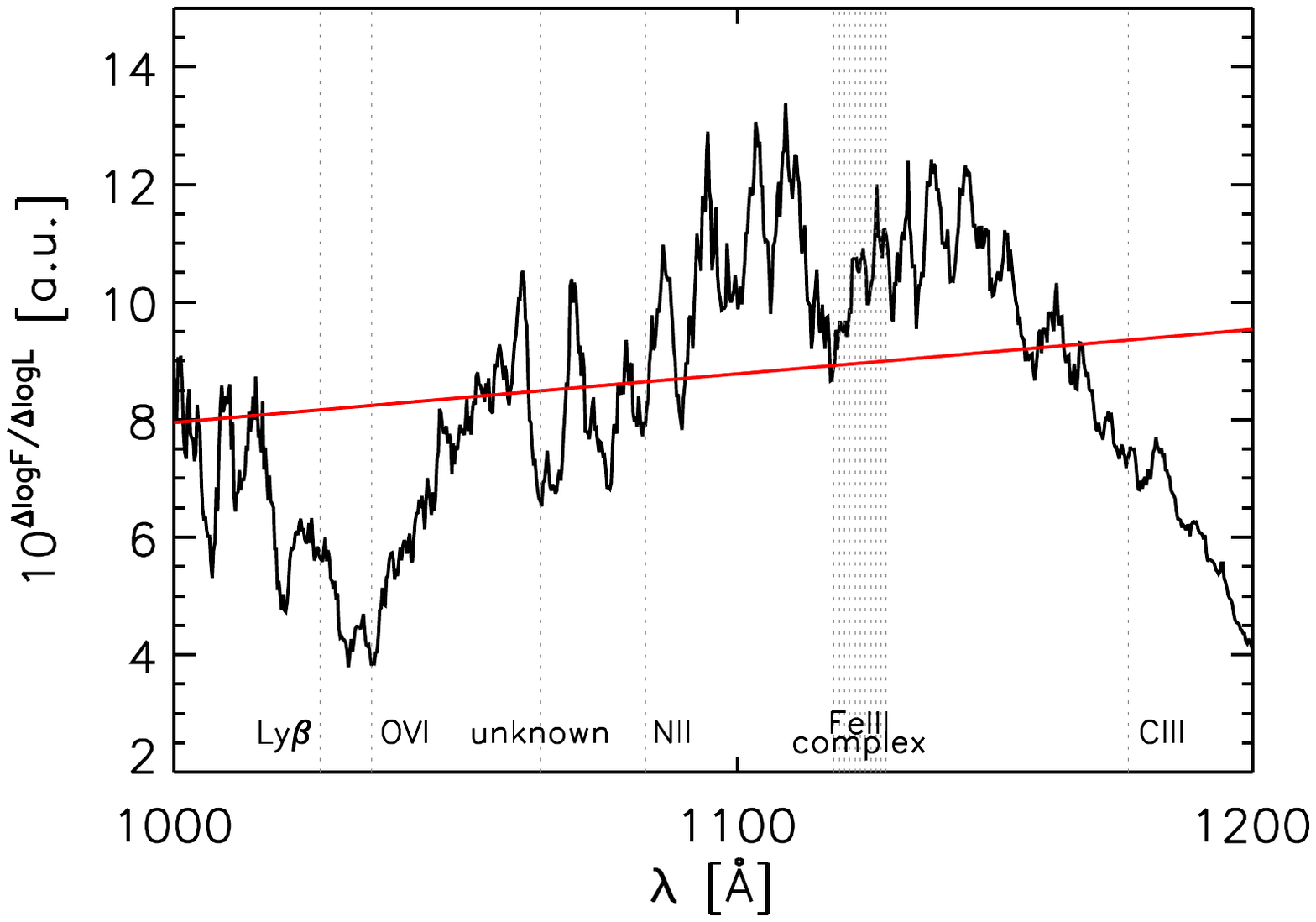} %, bb=75 365 550 725
                \includegraphics[bb=65 355 550 715, width=0.49\textwidth]{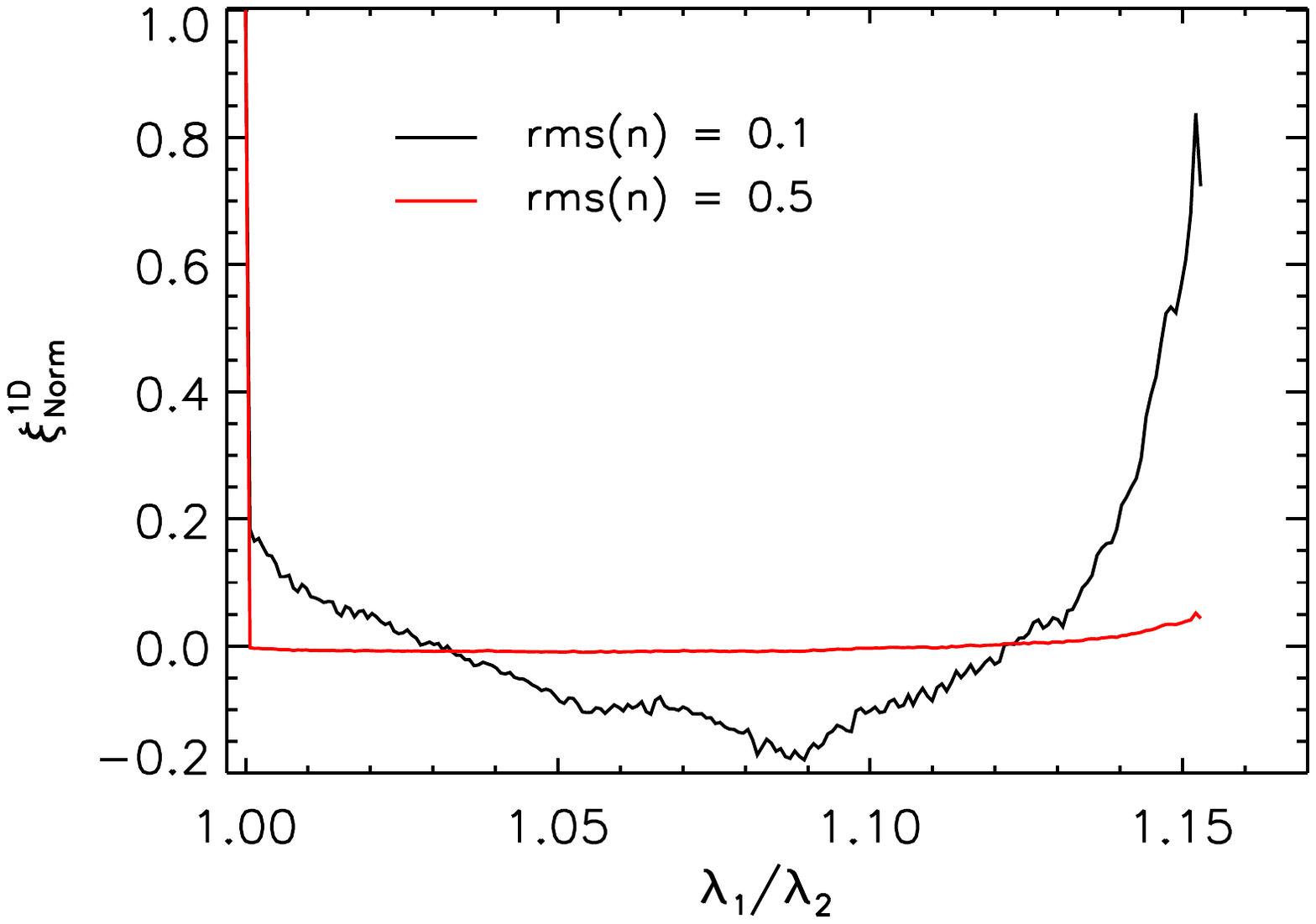} %, bb=75 365 550 725, bb=80 360 575 705
                \caption{{\bf Left:} The composite differential spectrum shown in Figure~\ref{fig:logdfdl} (black) represented as a fractional change in flux. The best linear fit to this signature (red) represents the model used in BOSS and eBOSS cosmology studies to describe quasar diversity. {\bf Right:} The normalized 1D correlation of two pixels in the same forest as a function of wavelength ratio ($\xi^{\mathrm{1D}}_{\mathrm{Norm}}=\xi^{1D}(\lambda_1/\lambda_2) / \sqrt{\xi^{1D}(\lambda_1)\xi^{1D}(\lambda_2)}$). This signature was made from spectra simulated according to Equation~\ref{eq:simspec} and the publicly available code ``picca'' (https://github.com/igmhub/picca).}
                \label{fig:autocorrelation}
        \end{center}
\end{figure*}

All of the differential spectra produced in this work have significant structure
over the rest-frame wavelength interval 1040~-- 1200~\AA. This is the region that is used in BOSS and eBOSS studies of clustering in
the absorption of quasar spectra due to the Lyman-$\alpha$ forest \citep{Bautista, Helion,1dforneutrinos}.
The left hand panel of Figure~\ref{fig:autocorrelation} shows the fractional change in flux associated with an order of magnitude
change in luminosity over this wavelength range.  For simplicity, we use only the differential spectrum in
Figure~\ref{fig:logdfdl}, although we note that the results would not change significantly if other models were used.
This signature of variability is one mode of the diversity that is present in the full sample of quasars in the BOSS and eBOSS Lyman-$\alpha$ forest studies.
Because the underlying continuum of quasars must be uncorrelated between lines of sight, this signal should not distort the measurements
of three-dimensional clustering. However, if not incorporated into the model for the quasar continuum, this diversity will
increase the measurement uncertainties of the flux-transmission field. The unmodeled signature in Figures~\ref{fig:logdfdl} and~\ref{fig:autocorrelation}
does have the potential to bias the cosmological interpretation in one-dimensional power spectrum measurements because the fluctuations
in the flux-transmission field are extracted solely from the line of sight of each quasar.

In determining the transmitted flux fraction for BOSS and eBOSS clustering studies, the continuum over the range
$1040 < \lambda_{\rm RF} < 1200$ \AA\ is modeled for each quasar as an average spectrum scaled by a term that is linear in $\log{\lambda}$
to account for diversity. The best linear fit to the differential spectrum is shown in red in the left hand panel of Figure~\ref{fig:autocorrelation}.
As one would expect, the linear fit is unable to capture the diversity associated with the intrinsic Baldwin Effect in the red tail of the Lyman-$\beta$(1025~\AA)/\textsc{O~vi}(1034~\AA) emission line blend and the blue tail of the Lyman-$\alpha$ emission at 1216~\AA.  In addition, the evolving spectral index over this wavelength range introduces a broadband residual from the linear fit.

To quantify the bias introduced to clustering analyses by incomplete modeling of this signature of diversity, we determine the normalized one-dimensional correlation function from a sample of 100,000 synthetic quasar spectra. These spectra are generated by applying the modes of variability presented in Figure~\ref{fig:logdfdl} to the mean quasar flux ($F_\mathrm{avg}$) found in \citet{Helion} as follows:
\begin{equation}
F_{\mathrm{syn}}=10^{ \left(\log{F_\mathrm{avg}} + m(\lambda) * \Delta \log{L_{\mathrm{bol,syn}}} \right)} + n.
\label{eq:simspec}
\end{equation}
Changes in luminosity ($\Delta \log{L_{\mathrm{bol,syn}}}$) for each synthetic quasar spectrum are randomly sampled from a Gaussian distribution with rms($\Delta \log{L_{\mathrm{bol,syn}}}$) = 0.1. The width of this distribution is determined from the distribution of luminosities about the steady state for the average quasar in the SDSS RM program. $m(\lambda) = \frac{\Delta \log{F}}{\Delta \log{L_{\mathrm{bol}}}}(\lambda)$ is the signature shown in Figure~\ref{fig:logdfdl}, and the measurement uncertainty is randomly sampled for each pixel according to the term $n$. The resulting 1D auto-correlation for these spectra is shown in the right panel of Figure~\ref{fig:autocorrelation} for both a uniform rms of $n=0.1$ (black; S/N per pixel $\sim$ 10) and a uniform rms of $n=0.5$ (red; S/N per pixel $\sim$ 2).

The normalized one-dimensional correlation function is somewhat complicated as it depends on the amplitude of the assumed measurement uncertainties.  Aside from the signature of measurement uncertainties, several trends are clear in the correlation function. First, the primary signature appears as broad-band correlation over the full forest. This signal can primarily be described by long-wavelength modes in the 1D power spectrum. Because the cosmology constraints derived from 1D power spectrum measurements rely on larger wavenumbers ($k>0.002~\mathrm{km \, s}^{-1}$), the majority of contamination from the continuum residuals is excluded from the analysis.  This broadband signal is expected to increase the noise on all scales in the measured three-dimensional correlation function. Second, the residuals from correlation between the Lyman-$\beta$(1025~\AA)/\textsc{O~vi}(1034~\AA) blend and Lyman-$\alpha$ emission produce a strong signal at a separation $\lambda_1/\lambda_2 = 1.15$. As with the broadband signal, this separation is too large to have a significant impact on cosmology derived from the one-dimensional power spectrum.  That said, this unmodeled signal likely introduces noise to the three-dimensional correlation function on scales larger than the BAO scale. Finally, there is no significant signal due to the other emission lines in the Lyman-$\alpha$ forest. The largest equivalent widths measured in a high signal-to-noise composite spectrum \citep{Harris2016} are due to emission from an unknown line at 1064~\AA, \textsc{N~ii} at 1083~\AA, and \textsc{C~iii} at 1175~\AA.  While there is evidence in the differential spectrum for blended emission from an Fe~\textsc{iii} complex (1118~-- 1128~\AA) and \textsc{N~ii} at at 1083~\AA, the signals are broad and diluted in the one dimensional correlation function. There is possible signature at a separation around $\lambda_1/\lambda_2 = 1.08$.  This could be due to the residuals in the Fe~\textsc{iii} complex correlated with both Lyman-$\alpha$ and Lyman-$\beta$(1025~\AA)/\textsc{O~vi}(1034~\AA) emission residuals.

Although there does not appear to be strong evidence for biased measurements of clustering due to the spectral diversity associated with quasar variability in the forest,
there remains potential to improve the cosmological impact of current samples.
If the continuum can be fit purely with data from unabsorbed quasar continuum at longer wavelengths \citep[e.g.][]{eilers17,davies18}, the predicted spectrum in the Lyman-$\alpha$
forest region will allow unbiased estimates of the underlying quasar continuum level and the flux-transmission field.  Such an improvement would allow
new cosmology constraints from the Lyman-$\alpha$ forest that are informed by the full shape of the power spectrum.
For example, an unbiased model of the quasar continuum level would allow new constraints on the spectral index of primordial density
fluctuations ($n_s$), the change in the spectral index with wavenumber ($d n_s/d {\rm ln} k$), and the effective number of neutrino species ($N_\mathrm{eff}$). In the remainder of this section,
we estimate the potential of the modes of diversity presented in Figure~\ref{fig:binned_dfdl} for predicting the Lyman-$\alpha$ continuum levels.

For the quasars in our sample that have complete coverage of the Lyman-$\alpha$ forest at $\lambda>1040$ \AA, we assume that the steady state model of each quasar is known and model only the variability.
For all epochs of all of these quasars, the median measurement error of the continuum level over the interval $1040 < \lambda_{\rm RF} < 1200$ \AA\ is 2.2\%.
Consistent with what was shown in Table~\ref{table:bins}, the continuum level over this wavelength range has an observed dispersion of 17.2\% due to the combination
of quasar variability, flux measurement uncertainties, and flux calibration errors.
We fit each epoch of every quasar at wavelengths $\lambda>1216$ \AA\ with the steady-state spectrum for that quasar and the differential spectrum scaled by a single free parameter.
The model corresponding to the minimum $\chi^2$ is assumed to represent the quasar over the full wavelength range.  We assess the flux residual over the
interval $1040 < \lambda_{\rm RF} < 1200$ \AA\ of each epoch of each quasar relative to its model.
When doing so for each of the five differential spectra found in Figure~\ref{fig:binned_dfdl}, we find that all models reduce the dispersion
by approximately a factor of two (Table~\ref{table:lyadispersion}).  
A simple model of only the steady-state quasar spectrum fit with a single scaling factor to the observed spectrum at each epoch produces
a reduction in the dispersion comparable to the differential spectra models.  In this case, where no physically-motivated variability signature is assumed, the dispersion in the
Lyman-$\alpha$ continuum levels is reduced to 10.5\%. 

While our differential signature of variability is a significant improvement over the unmodeled
Lyman-$\alpha$ forest continuum, there is still room for another factor of three improvement on the dispersion.  
Flux calibration errors appear to dominate the remaining dispersion and limit the potential to improve the predictive power of the unabsorbed
quasar continuum much beyond what is shown in Table~\ref{table:lyadispersion}.
The flux calibration errors were computed from the standard star spectra that were used to compute the flux calibration for each epoch of the SDSS~RM program.
For all epochs of each star, we compute the total flux in a bandpass that corresponds to the rest-frame wavelength ranges presented in Table~\ref{table:bins}
for a quasar at redshift $z=2.62$.
We find that the dispersion in stellar fluxes over the observer frame bandpass corresponding to $\lambda_{\rm eff,Lum}=1100$ \AA\ is 9.1\%.  The correlation between this bandpass
approximating the Lyman-$\alpha$ forest continuum and the widest bandpass ($g+r$) that best approximates the unabsorbed quasar spectrum is only 27\%.
The result implies that the information
at longer wavelengths cannot reduce the dispersion much below 9.1\% for a quasar at $z=2.62$.  The dispersion in flux calibration decreases with increase in
wavelength, so the predictive power should increase somewhat for the Lyman-$\alpha$ forest continuum in quasars at higher redshift.
We conclude that the models of variability do improve estimates of the intrinsic quasar continuum in the Lyman-$\alpha$ forest, but
the true potential cannot be reached without better flux calibration.

\begin{table}%[!t] 
        \begin{center}
                \caption{
                \label{table:lyadispersion}
                Dispersion in estimates of the Lyman-$\alpha$ forest level as a function of assumed model.}
                \begin{tabular}{c c c }
                \hline \hline
                Model of & $\lambda_{\rm eff,Lum}$ & Fractional Dispersion \\
                Variability & & After Fit \\
                \hline
                $F(\lambda)_{\mathrm{steady-state}}$ & N/A & 0.172 \\
                $A * F(\lambda)_{\mathrm{steady-state}}$ & N/A & 0.105 \\
                $F(\lambda)_{\mathrm{steady-state}}*10^{A*m(\lambda)}$ & 1100 \AA\ & 0.080 \\
                $F(\lambda)_{\mathrm{steady-state}}*10^{A*m(\lambda)}$ & 1400 \AA\ & 0.077 \\
                $F(\lambda)_{\mathrm{steady-state}}*10^{A*m(\lambda)}$ & 1740 \AA\ & 0.088 \\
                $F(\lambda)_{\mathrm{steady-state}}*10^{A*m(\lambda)}$ & 2100 \AA\ & 0.098 \\
                $F(\lambda)_{\mathrm{steady-state}}*10^{A*m(\lambda)}$ & $g+r$ & 0.088 \\
                \hline
                \end{tabular}
        \end{center}
\end{table}

\newpage
\section{\textbf{Conclusion}}\label{sec:con}

In this work, we use 340 quasars ($1.62 < z < 3.30$) and a total of 16,421 spectra from the Sloan Digital Sky Survey Reverberation Mapping program to quantify the spectral variability of quasars. Simple linear relationships of $\log{(\textsc{C~iv} W_\lambda)}$ and $\alpha_\lambda$ with $\log{L_{\mathrm{bol}}}$ in our spectra reproduce the previously established Baldwin and bluer-means-brighter effects, respectively.

We further investigated these trends with a composite differential spectrum which evaluates the logarithmic changes in flux as a function of the change in the log of luminosity. The composite differential spectrum is shown in Figure~\ref{fig:logdfdl}.  It follows a power law redward of about 1700~\AA\ and reveals detailed trends in continuum-emission line correlations and color evolution as a function of discrete wavelength, further substantiating the Baldwin and bluer-means-brighter trends. The most surprising characteristic of our differential signature is a break in the differential power law blueward of 1700~\AA.

We explored this break in the differential power law by recreating the differential spectrum using different definitions of luminosity.  The position and strength of the break
vary depending on which bandpass is used to assess the change in quasar brightness.  Specifically, the differential spectra tend to peak near the effective
wavelength of the luminosity window and the amplitude of the differential spectra tend to drop off more quickly at larger separations in wavelength.
The correlation coefficients in the luminosity estimates reinforce the observation that the fluctuations in flux grow decorrelated at large separations in wavelength.
The evidence implies that peak fluctuations tend to be localized over a relatively narrow wavelength range and have limited power in predicting
the spectral response at other wavelengths.

We then argued that the observed variability can be explained by either the inhomogeneous or slim accretion disk models. The stochastic, uncorrelated nature of variability
arises naturally from inhomogeneous accretion disk structure, as shown in a series of simulated spectra.
The behavior could alternatively be a product of slim accretion disk structure.  In this case, X-ray and FUV flux would be suppressed by an optically-thick shielding gas.
A change in the shielding gas scale height and radial extent with increasing Eddington effiency, combined with a delayed response in the accretion disk to reprocessing of
incident flux, could explain the observed behavior.

Finally, we used our composite differential spectrum to quantify the potential in introducing new templates to the modeling of the
quasar continuum in cosmological Lyman-$\alpha$ forest analyses.  We first explore the bias introduced by linear models of diversity in quasar continua.
We show that incorrect modeling of quasar diversity around the Lyman-$\alpha$ forest does introduce spurious structure, but this structure appears on scales that are not
used in the cosmological studies of the one-dimensional power spectrum.  The structure is also likely to increase the noise, rather than introduce significant biases,
to studies of large-scale structure in the Lyman-$\alpha$ forest.  This result is consistent with earlier cosmological analyses \citep[e.g. ][]{slosar11a}.
When applying the differential spectra as models to predict the Lyman-$\alpha$ forest continuum, we find that the dispersion due to variability
can be reduced from 17.2\% to as little at 7.7\%.  We find that simpler models with no accounting for the signature of variability are able to reduce
the dispersion to 10.5\%.
Given the variance in Lyman-$\alpha$ continuum levels due to flux calibration errors,
further work to improve the accuracy of relative calibration is required to improve the power of quasar spectral models to directly predict the level
of the quasar continuum in the Lyman-$\alpha$ forest.  The data reduction pipeline for eBOSS is nearly finalized and unlikely to produce significant
improvements in calibration, but potential exists for improved spectrophotometry in the Dark Energy Spectroscopic Instrument \citep[DESI;][]{DESI}.

\acknowledgements

Funding for the Sloan Digital Sky Survey IV has been provided by the Alfred P. Sloan Foundation, the U.S. Department of Energy Office of Science, and the Participating Institutions. SDSS-IV acknowledges support and resources from the Center for High-Performance computing at the University of Utah. The SDSS web site is www.sdss.org. SDSS-IV is managed by the Astrophysical Research Consortium for the Participating Institutions of the SDSS Collaboration including the Brazilian Participation Group, the Carnegie Institution for Science, Carnegie Mellon University, the Chilean Participation Group, the French Participation Group, Harvard-Smithsonian Center for Astrophysics, Instituto de Astrof\'isica de Canarias, The Johns Hopkins University, Kavli Institute for the Physics and Mathematics of the Universe (IPMU) / University of Tokyo, Lawrence Berkeley National Laboratory, Leibniz Institut f\"ur Astrophysik Potsdam (AIP), Max-Planck-Institut f\"ur Astronomie (MPIA Heidelberg), Max-Planck-Institut f\"ur Astrophysik (MPA Garching), Max-Planck-Institut f\"ur Extraterrestrische Physik (MPE), National Astronomical Observatories of China, New Mexico State University, New York University, University of Notre Dame, Observat\'ario Nacional / MCTI, The Ohio State University, Pennsylvania State University, Shanghai Astronomical Observatory, United Kingdom Participation Group, Universidad Nacional Aut\'onoma de M\'exico, University of Arizona, University of Colorado Boulder, University of Oxford, University of Portsmouth, University of Utah, University of Virginia, University of Washington, University of Wisconsin, Vanderbilt University, and Yale University.

The work of KD, HdMdB, and MV was supported in
part by U.S. Department of Energy, Office of Science,
Office of High Energy Physics, under Award Number DESC0009959.
We thank Nicol{\'a}s~G.~Busca for input regarding the {\tt picca} software repository and for helpful comments on continuum fitting for Lyman-$\alpha$ forest studies.
We thank Jonathan Trump, Yue Shen, Pat Hall, and Keith Horne for sharing preliminary results from studies of the rms spectra from this quasar sample.
Finally, we thank Gordon Richards for feedback on the text of the paper.
 
\bibliographystyle{mnras} 
\bibliography{bibliography} 

\end{document}